\def\aj{\ {AJ}\ }

\def\apjl{\ {ApJL}\ }

\def\mnras{\ {MNRAS}\ }
\def\nat{\ {Nat}\ }

\def\pasj{\ {Publ. Astr. Soc. Japan}\ }

\documentclass[conference]{IEEEtran}
\usepackage[pdftex]{graphicx}

\usepackage{url}
\input ./abbreviations.inp

\usepackage{graphicx}
\usepackage{psfig}

\usepackage{ifthen}
\newboolean{SimonsSetup}
\setboolean{SimonsSetup}{true}

\def\apgt{\ {\raise-.5ex\hbox{$\buildrel>\over\sim$}}\ } 
\def\aplt{\ {\raise-.5ex\hbox{$\buildrel<\over\sim$}}\ } 
\def\lt{\ {\raise-.5ex\hbox{$\buildrel>$}}\ } 
\def\gt{\ {\raise-.5ex\hbox{$\buildrel<$}}\ }

\def\half{\ensuremath{1 \over 2}}

\def\Msun{\ensuremath{{\rm M}_{\odot}}}

\begin{document}

\title{Computational Gravitational Dynamics with Modern Numerical Accelerators}

\author{Simon Portegies Zwart, 
        Jeroen B\'edorf \\
\IEEEauthorblockA{Sterrewacht Leiden, Leiden University, 
                P.O. Box 9513, 2300 RA Leiden, The Netherlands}
        }


\maketitle

\begin{abstract}

We review the recent optimizations of gravitational $N$-body kernels
for running them on graphics processing units (GPUs), on single hosts
and massive parallel platforms. For each of the two main $N$-body
techniques, direct summation and tree-codes, we discuss the
optimization strategy, which is different for each algorithm.  Because
both the accuracy as well as the performance characteristics differ,
hybridizing the two algorithms is essential when simulating a large
$N$-body system with high-density structures containing few particles,
and with low-density structures containing many particles. We
demonstrate how this can be realized by splitting the underlying
Hamiltonian, and we subsequently demonstrate the efficiency and
accuracy of the hybrid code by simulating a group of 11 merging
galaxies with massive black holes in the nuclei.

\end{abstract}


\IEEEpeerreviewmaketitle

\section{Introduction} 

\begin{figure}
\centering
\begin{minipage}[t]{\textwidth}
\includegraphics[width=\textwidth,height=90pt]{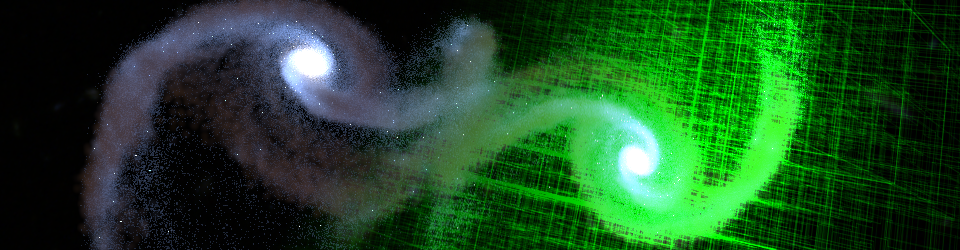}
\caption{Rendering (left) and computer representation (tree-code in green) 
of a collision between two galaxies, run on the Titan supercomputer.
\label{fig:rendering}}
\end{minipage}
\end{figure}


The first recordings of theoretical astronomy date back to the
Egyptian 18th dynastic (1550-1292 BC) calculations by pharao
Hatshepsut's architect Senmut; part of the murals found in his tomb at
Del el-Bahri is exhibited at \url{http://www.metmuseum.org}.  Such
calculations enable astronomers to recognize structure and describe
patterns in the heavens; a branch of astronomy that is still relevant
for classification and charting.  Today's astronomers hardly use clay
tablets or stilae, but the instruments with which observations are
conducted have been growing gradually. 

The size of telescopes has increased from the $1\,{\rm cm}^2$
refractor that Lippershey build in 1608, to the $\sim 110\,{\rm m}^2$
collecting area of the Gran Telescopio Canarias.  This 20-fold
doubling of collecting area has been achieved in the last 400
years. Digital computers were only introduced at the end of the 1940s
starting with a computational speed of about 100 modern floating point
operations per second (or flops, ENIAC could perform about 360
multiplications in 6 decimal places per second) to about $3 \cdot
10^{16}$\,flops today; a 44-fold doubling in raw performance in only
65 years.  Astronomers have therefore 
\vfill
\vspace*{105pt}
\noindent
grown adiabatically in the improvement of their instruments, whereas
computer scientists have experienced an explosive evolution.

This revolution in the availability of digital computers is still
ongoing and has led to an entirely new branch of research in which
facilities are not on high mountain tops on the Canary islands, but in
the room next door.  Astronomers realized quickly that they could use
computers to archive, process, analyze and mine the copious amounts of
data taken by observing campaigns.  The biggest impact in the way
astronomers pursued their scientific questions however, has been by
means of simulation.

The fundamental complexity of astronomical research lends itself
ideally for computation, because it is characterized by the enormous
temporal and spatial scales, complex non-linear processes and the
extremes of space. With the introduction of the digital computer it
suddenly became possible to study processes in the intergalactic
vacuum, hot plasma's at stellar surfaces, billion-year time-scale
processes and black hole physics; none of which can be studied in 
Earth-based laboratories.

The largest remaining limitation in studying the universe by means of
computation are introduced by the software, and in particular our
limited understanding of the algorithms for resolving the wide range
of temporal and spatial scales, and for solving coupled fundamental
physical processes.

Here we report on our experience in designing new algorithms for
solving some of the software-related problems using recent hardware
developments of attached accelerators. We limit ourselves to
gravitational dynamics, and in particular to the gravitational
$N$-body problem \cite{Aarseth2003}, because here the developments
have been quite pronounced and the application is sufficiently general
that the algorithms can be generalized to other research fields.

\section{The separation of direct and hierarchical methods}

Gravitational $N$-body dynamics poses an excellent research field for
computing, because the physics based on Newton's law of gravity can be
calculated from first principles and has not changed (much) since
1687. The consequences of this long-range energy-conserving force are
pronounced and extremely difficult to study analytically, leaving the
computer as the only remaining alternative for research.

Shortly after the first $N$-body codes emerged, software development
took an interesting direction in simulating more extended systems with
more self-gravitating elements (i.e.\, particles) at the cost of a
lower spatial and temporal resolution.  This separation was initiated
by the development of the tree-algorithm \cite{1986Natur.324..446B},
which introduced a collision-less approximation scheme based on the
natural hierarchy consequented by the long-range characteristic of the
interaction.

From this moment the development of $N$-body simulation codes branches
in two quite distinct tracks: these are the direct force evaluations
which scale $\propto N^2$ and approximate force evaluators which scale
$\propto N \log N$ or better. The forces of the former are generally
used in combination with high $\ge 4$th order integrators with
individual time stepping \cite{2008NewA...13..498N} whereas the latter
are generally combined with phase-space volume-preserving symplectic
integrators\footnote{One can read more about symplectic integrators
at \url{http://en.wikipedia.org/wiki/Symplectic_integrator}.} \cite{1991AJ....102.1528W}. (Although
there are production codes, in particular for simulating planetary
systems, in which direct force evaluations are combined with
symplectic integrators.)  After this rather strict separation in
philosophy we will discuss here how advances in software and hardware
have recently led to the reunion of both algorithms.  But before we
discuss the reunion of algorithms, we discuss the revolution in both
field separately.

\section{Simulating Collisional systems with direct force evaluation methods}
\label{Sect:direct-integration}

Direct $N$-body calculations are generally adopted for studying the
dynamical evolution of collisional systems over a relaxation time
scale. Areas of research include stability studies of planetary
systems, the evolution of star clusters and galactic nuclei with
massive black holes.

Enormous advances in software have been achieved by the introduction
of individual time-steps, the Ahmad-Cohen neighbor scheme and a rich
plethora of regularization techniques.  Each of these however, are
hard to parallelize.  The introduction of block time steps partially
solved this problem and opened the way to separate the computational
part of the calculation from all-to-all communication, in so called
$i$-particle parallelization. The introduction of block-time steps has
motivated the development of the GRAPE-family of special purpose
computers, and eventual led to to high-performance gravitational
$N$-body simulations using attached accelerator hardware, like
graphics processing units (GPUs) \cite{2007NewA...12..641P}.  See the
reviews \cite{1998sssp.book.....M,Aarseth2003}.

The largest simulations conducted using direct $N$-body methods are
approaching a million objects.  But this is still a small number
compared to the Solar system which is composed of one star, 8 planets,
166 moons and several million planetesimals, or the Milky Way Galaxy,
which is composed of $\sim 100$~billion stars, each of which may be
accompanied by a million-body planetary system.

\subsection{Optimizations for the GPU}

For convenience we separate the particles in the stellar system in $j$
particles, which exert a force, and $i$ particles that receive a
force.  To solve the $N$-body problem the forces exerted by the
$j$-particles onto the $i$-particles have to be computed. The
particles in subset $j$ can either belong to the same or a completely
different set as the $i$-particles. In a worst case the algorithms
scale as $N_j N_i \rightarrow N^2$, but because the forces calculated
on the $i$-particles are independent the algorithm is embarrassingly
parallel for $p=N_i$ cores. However, by design in the individual time
step method $N_i \ll N_j$, because only the particles are updated that
require a force update at a certain 
time \cite{1998sssp.book.....M,Aarseth2003}.

On a GPU the parallelization can be exploited by launching a separate
compute thread for each $i$ particle \cite{2007NewA...12..641P} This
is efficient only for $N_i \apgt
10^4$ \cite{1998sssp.book.....M,Aarseth2003} to warrant the saturation
of all compute threads.  For a smaller number of $i$ particles the
number of active compute threads is not sufficient to hide the memory
and instruction latencies.  Future devices may require an even larger
number of running compute threads to reach peak performance, causing
$N_i$ to be even larger before the device is saturated.  Adjusting the
number of $i$ particles to keep parallel efficiency is not ideal.  In
an alternative approach, followed by \cite{2004PASJ...56..521M}, one
can parallelize the $j$-particles, while fixing the number of $i$
particles over which we concurrently integrate.  We split the
$j$-particles in subsets which form the input against which a block of
$i$ particles is integrated.  The number of $j$-particles per block
then increases for smaller $N_i$, making the algorithm efficient even
for relatively small $N_i$. This method can fully utilize the GPU
performance when the product of $N_i$ and the number of subsets in
which the $j$ particles exceed the number of compute threads.

Earlier GPU hardware lacked support for double precision arithmetic,
but this did not necessarily pose a problem for most common
algorithms, even with a fourth-order Hermite
integrator \cite{1998sssp.book.....M}.  The largest error in these
cases is made in the calculation of the inter-particle distance, but
this can be solved by emulating double precision on single-precision
hardware.  The integration of the orbits of objects with an extreme
mass-ratio, like planetesimals around super-massive black holes,
remains hard to achieve without IEEE-754 double precision arithmetic.
The latest GPUs support double precision and therefore enable such
research.

Running high-performance $N$-body simulations on multiple GPUs has
become common practice, even though distributing the calculation of
several nodes proves less effective due to the limited bandwidth.
Latency is not much of a problem because it turns out to be relatively
easy to hide most of the latency in other operations.  Multi-GPU
parallelization is achieved by distributing the $j$ particles over the
various GPUs using a round-robin method.  This reduces the memory
usage, transfer time and the time required to execute the optional
prediction step on the $j$ particles.  In this scheme the first kernel
computes the partial forces of our parallelization over the $j$
particles, and the second kernel sums these.

\begin{figure}[t]
\begin{center}
\includegraphics[width=0.5\textwidth]{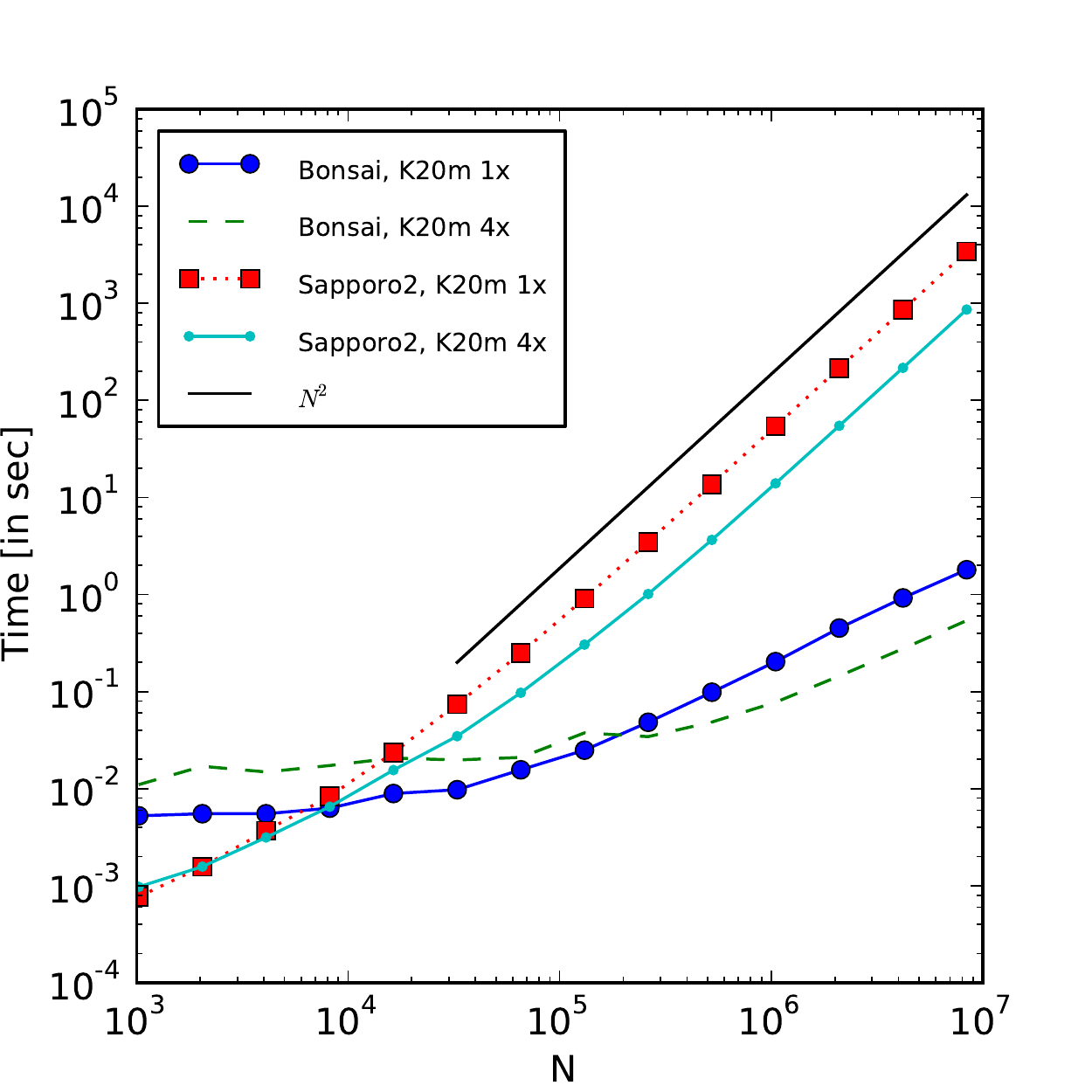}
\end{center}
\caption{Wall-clock time as a function of the number of particles in a
         direct code (upper curves) and a tree code (bottom curves
         using $\theta =0.4$).
\label{fig:Performance}}
\end{figure}

The introduction of efficient atomic operators in the latest NVIDIA
GPUs made it possible to combine the calculation of the partial forces
of the $j$ particles with the actual summation over all forces.  This
combined operation increases performance and reduces the complexity of
maintaining the compute kernels. The above described method is
implemented in {\tt Sapporo2} (a GPU library for solving the
gravitational $N$-body problem)~\cite{2009NewA...14..630G} where the amount of
work is arbitrarily split up in many independent computation blocks
and automatically scale to future architectures.  In
Fig.~\ref{fig:Performance} we present the performance characteristic
of the {\tt Sapporo2} library running a single GPU, and four GPUs for
$10^3$ to $10^7$ particles.

\section{Simulating collisionless systems with hierarchical domain-decomposition methods}
\label{Sect:hierarchical}

The Barnes-Hut tree algorithm
\cite{1986Natur.324..446B} 
turned into a classic shortly after its introduction.  In this
algorithm the distribution of particles is recursively partitioned
into octants until the number of particles in an octant is smaller
than a critical value. Once a tree is built and its multipole moments
are computed, the code proceeds by calculating the forces.  For this,
we adopt a geometric angle criterion, called the multipole acceptance
criterion, $\theta$, which purpose is to decide whether or not the
substructure in distant octants can be used as a whole.  For finite
$\theta$ however, sufficiently distant octants from a target particle
can be used as a whole, and the partial forces from the constituent
particles are calculated via a multipole expansion approximation.  The
time complexity for calculating the force between all particles in the
entire system reduces with the tree-code to $\propto N\log(N)$.

The largest simulations conducted using tree codes are approaching 100
billion objects.  This number is sufficient so simulate the entire
Milky Way Galaxy on a star-by-star basis. The accuracy, however, is
insufficient to simulate the intricate dynamics of the dense stellar
systems we discussed in \S\,\ref{Sect:direct-integration}.

\subsection{Implementation}\label{Sect:GPUoptimizations}

In the Barnes-Hut algorithm we can distinguish three fundamental
parts: the construction of the tree, computation of quadrupole
moments, and the tree-walk in which inter-particle gravitational
forces are computed.  Traditionally the tree-walk and gravity
computation take up the largest part of the computation time. However,
if this part would be optimized with the help of accelerators, like
the GPU, other parts of the algorithm would become bottlenecks. Either
because their relative contribution in the execution time becomes
larger or because of the requirement to send data back and forth
between the accelerator and the host CPU.

One way to eliminate these bottlenecks and data-transfers is by
porting all the computational parts of the tree-code to the GPU. This
is the approach taken in {\tt Bonsai} \cite{2012JCoPh.231.2825B}.  In
this code the tree-walk and force computations are assimilated into a
single GPU kernel which allows for excellent computational efficiency
by not wasting GPU bandwidth to store particle interaction lists into
memory.  Instead, the interaction lists are stored in registers and
evaluated on-the-fly during the tree-walk, therefore delivering a
performance in excess of 1.7 Tflops on a single
K20X \cite{2012JCoPh.231.2825B}.

It is considerably more efficient to walk the tree on a GPU by a group
of spatially nearby particles rather than by one particle at a
time \cite{1986Natur.324..446B}. These nearby particles have similar
paths through the tree, and therefore similar interaction lists; by
building an interaction list that is valid everywhere within the
group, one can reduce the number of tree-walks and make each of them
efficient via thread cooperation inside a thread-block.  The grouping
in \cite{1986Natur.324..446B} is based on the underlying
tree-structure, such that tree-cells with the number of particles
below a certain number  are used as a
group. However, due to the geometric nature of space partitioning by
octrees, the average number of particles in such a group was much
smaller than 64, which wastes compute resources.  This is solved by
sorting the particles into a Peano-Hilbert space filling curve
(PH-SFC)~\cite{citeulike:2861104} and splitting it into groups of 64
particles. The final criterion is to enforce a maximal geometric size
of a group: if a group exceeds this size it is recursively split
further into smaller groups.  An extra benefit of using the PH-SFC is
its ability to construct the tree-structure by splitting the SFC in
distinct nodes.

\subsection{Parallelization}\label{Sect:ParallelOptimizations}

While accelerators greatly improve the efficiency per node, massive
parallel computing with GPUs has become a challenging problem. One
crucial bottleneck is the relatively slow communication between the
GPU and its host, which directly limits the overall performance.
However, due to the long-range nature of Newton's universal law of
gravitation, the computation of mutual forces is by definition an
all-to-all operation. Which requires that data has to be exchanged
between all the nodes and therefore again requires communication
between the CPU and GPU and communication between the different CPUs,
which is even slower.

To maintain single-GPU efficiency when scaled to many GPUs requires
both the minimization of the amount of data traffic between different
GPUs, and hiding the communication steps behind computations. This is
realized by carefully selecting, combining and modifying different
well-known parallelization strategies.  In the following paragraphs we
describe this parallelization strategy.

\subsubsection{Domain Decomposition}

Each GPU computes its local domain boundaries, and the CPUs determine
global domain boundaries which are used for mapping particle
coordinates into corresponding PH keys~\cite{citeulike:2861104}, the
host subsequently gathers a sample of PH-keys from the remote
processes and combines these into a global PH-SFC. This SFC is cut
into $p$ equal pieces and the beginning and ending PH key determines
the sub-domains of the global domain, which are broadcast to all
processes.  The resulting domain boundaries will not be rectangular
but have fractal boundaries, which makes it hard to select particles
and nodes that are required on remote nodes.  As a
consequence, common SFC-based codes \cite{169640} generate multiple
communication steps during the tree-walk.  An alternative method is
the Local Essential Tree (LET) method.  In this method each process
uses the boundaries of the sub-domains to determine which part of its
local data will be required by a remote process. This part is called
the LET structure. After a process has received all the required LET
structures, they are merged into the local tree to compute the
gravitational forces.

This LET approach requires the least amount of communication and is
therefore preferred in a practical implementation, like in {\tt
Bonsai}. Here the LET method is uniquely combined with the SFC domain
decomposition which guarantees that sub-domain boundaries are
tree-branches of a hypothetical global octree. This allows for
skipping the merging of the imported structures into the local-tree,
and process them separately as soon as they arrive, therewith
effectively hiding communication behind computations.

\subsubsection{Computing the gravity}

To compute forces on local particles, a target process requires
communication with all other processes. In {\tt Bonsai} this is
realized by forcing remote processes to send the required particle and
cell data (via the LETs) to the target process. While the GPU on this
target process is busy computing forces from the local particles, the
CPU prepares particle data for export, as well as sending and
receiving data.

The preparation of particle data for export to remote processes is
both floating point and memory bandwidth intensive, both must overlap
with the communication between the processes. This is achieved by
multi-threading, in which each MPI process is split into three
thread-groups: one is responsible for communication (the communication
thread), one drives the GPU (the driver), and the rest are preparing
LET structures.  By disconnecting the various pieces of work it is
possible to let them overlap and keep force feeding the GPU with
computation work while other parts of the CPU take care of handling
the relatively slow operations related to inter-node communication.

In order to scale to thousands of compute nodes there is usually the
requirement that one needs to have trillions of particles that have to
be integrated.  The here discussed implementation of {\tt Bonsai}
allows for efficient scaling to thousands of nodes on the Titan
supercomputer when using a relatively modest amount of
particles.  

In Fig.\,\ref{fig:Performance} we present the performance of the
direct $N$-body and tree-code compute kernels both running on an NVIDIA
K20m GPU. The $N^2$ scaling of the direct code is clearly
distinguishable from the $N\log (N)$ scaling of the tree-code.
Increasing the number of GPUs in the calculations reduces the
wall-clock time but does not change the scaling characteristics, but
only changes the offset of the duration.

\section{The multi-scale approach}
\label{Sect:reunion}

The basis for the gravitational $N$-body problem is an integrable
Hamiltonian. The natural separation in collisional
(see \S\,\ref{Sect:direct-integration}) and collisionless
(see \S\,\ref{Sect:hierarchical}) domains can therefore also be
reflected in the Hamiltonian of the form: $H_{\rm regular} + H_{\rm
irregular}$, which can be solved numerically.  This operator splitting
approach has been demonstrated to work effectively and efficiently
by \cite{2007PASJ...59.1095F}, who adopted the Verlet-leapfrog
algorithm to combine two independent gravitational $N$-body solvers.  In
this case we can use the direct $N$-body code
(\S\,\ref{Sect:direct-integration}) and the tree-code
(\S\,\ref{Sect:hierarchical}) for those physical domains for which
they are most suited; the direct $N$-body code for simulating the
collisional environment and the tree code for simulating the
collisionless system. Such a hybrid numerical solver is ideally suited
for simulating planetary systems in a star cluster or the dynamics of
supermassive black holes in galactic nuclei.  In the next section we
will demonstrate the working of this hybrid approach on a problem in
which we allow 11 galaxies with black holes in their cores to merge.

In the scheme, the Hamiltonian of the entire system is divided into
two parts:
\begin{equation}
        H=H_A+H_B,
\end{equation}
where $H_A$ is the potential energy of the gravitational interactions
between galaxy particles and the stars in the sphere of influence of
the supermassive black hole in the galactic nuclei ($W_{g−c}$):
\begin{equation}
        H_A=W_{g−c},
\end{equation}
and $H_B$ is the sum of the total kinetic energy of all particles
($K_g+K_c$) and the potential energy of the galactic nuclei with
black hole ($W_c$) and the galaxy ($W_g$)
\begin{equation}
        H_B = K_g + W_g + K_c+W_c \equiv H_g + H_c.
\end{equation}
The time evolution of any quantity $f$ under this Hamiltonian can be
approximated (because we truncated the solution to second-order) as:
\begin{equation}
f'(t+\Delta t) \sim e^{\half \Delta t A} e^{\Delta t B} e^{\half \Delta t A} f(t),
\end{equation}
which represents a symplectic split in the Hamiltonian.  Here the
operators $A$ and $B$ are $A_f \equiv \{f,H_A\}, B_f \equiv
\{f,H_B\}$, and $\{.,.\}$ denotes the Poisson bracket. 
The evolution operator $e^{\Delta t B}$ splits into two independent
parts because $H_B$ consists of two independent parts without a cross
term. This embodies the second-order leapfrog algorithm, the time
evolution of which can be implemented as a kick---drift---kick scheme.

\section{Simulating galaxy mergers with supermassive black holes}
\label{Sect:simulations}

The hybridization described in \S\,\ref{Sect:reunion} is of an essence
if one is impatient for the results, but cannot afford to have an
approximate solution for all orbital integrations.  Simulating a one
billion particle galaxy-merger simulation easily takes several
decennia with an $N^2$ method, but only a few weeks with a tree
algorithm, whereas our implementation of the tree algorithm is
insufficiently accurate to resolve the intricate dynamical encounters
between the black holes (see Fig.\,\ref{fig:BHOrbits}). We therefore
integrate only the few thousand stars near each black hole using the
direct $N^2$ algorithm and all the other stars with the tree-code; in
this way we benefit from the advantages of both algorithms.

We apply our hybrid numerical scheme running on GPU accelerated
clusters using {\tt Bonsai} \cite{2012JCoPh.231.2825B} for the
tree-code and the 4th order Hermite predictor-corrector code called
{\tt ph4} as the direct $N$-body code. Both codes are incorporated in
the Astronomical Multipurpose Software environment,
AMUSE~\cite{PortegiesZwart2013456}.

\begin{figure}[t]
\begin{center}
\includegraphics[width=0.5\textwidth]{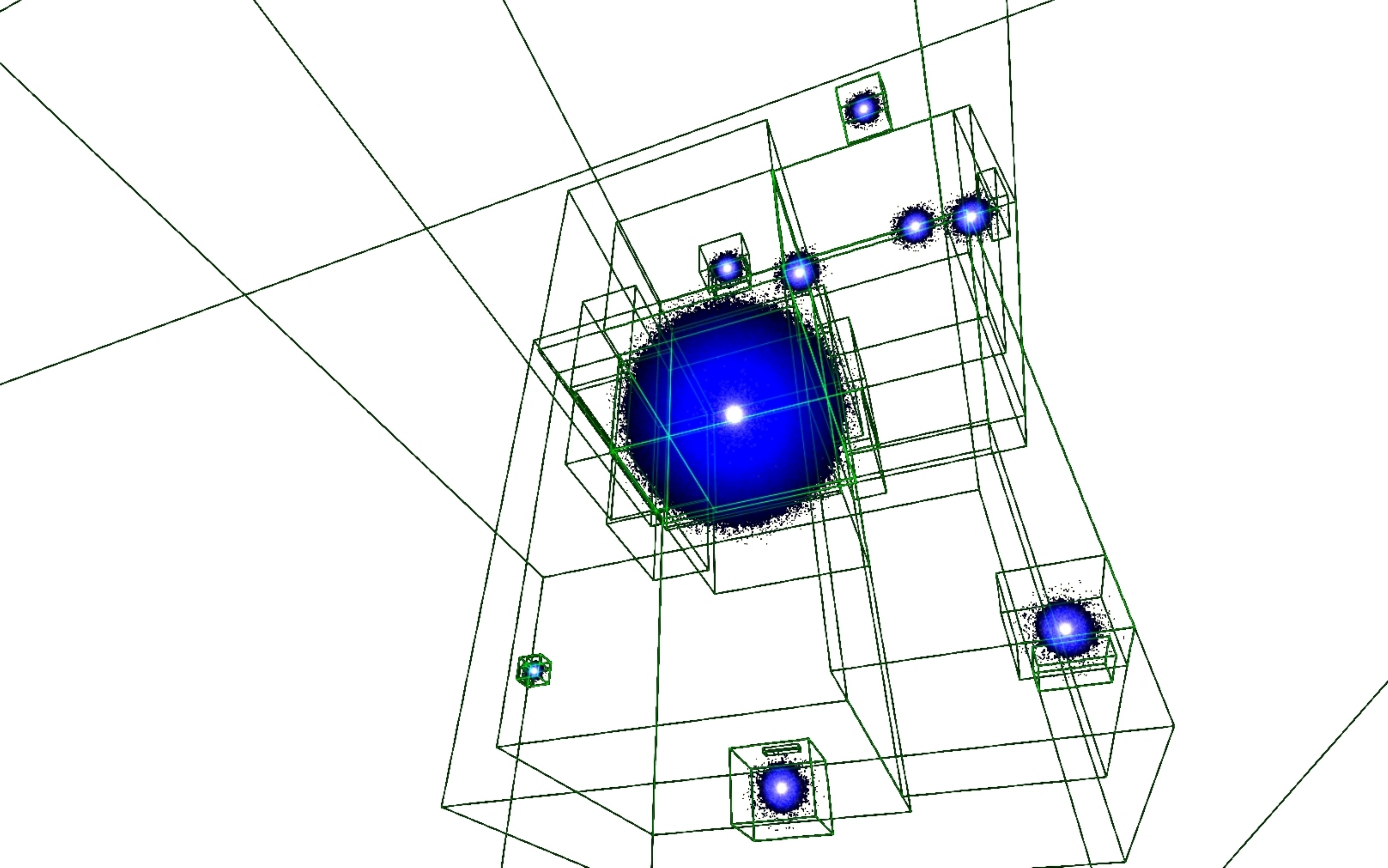}
\end{center}
\caption{Wide angle view of the initial realization of the simulated 
eleven galaxies (8 of the children and the parent are visible in blue)
and the outer most structure of the tree code (the green lines
indicate the tree-code boundaries). The 10 minor galaxies reside
initially within a distance of 500,000\,parsec.
\label{fig:t0}}
\end{figure}

\begin{figure}[t]
\begin{center}
\includegraphics[width=0.5\textwidth]{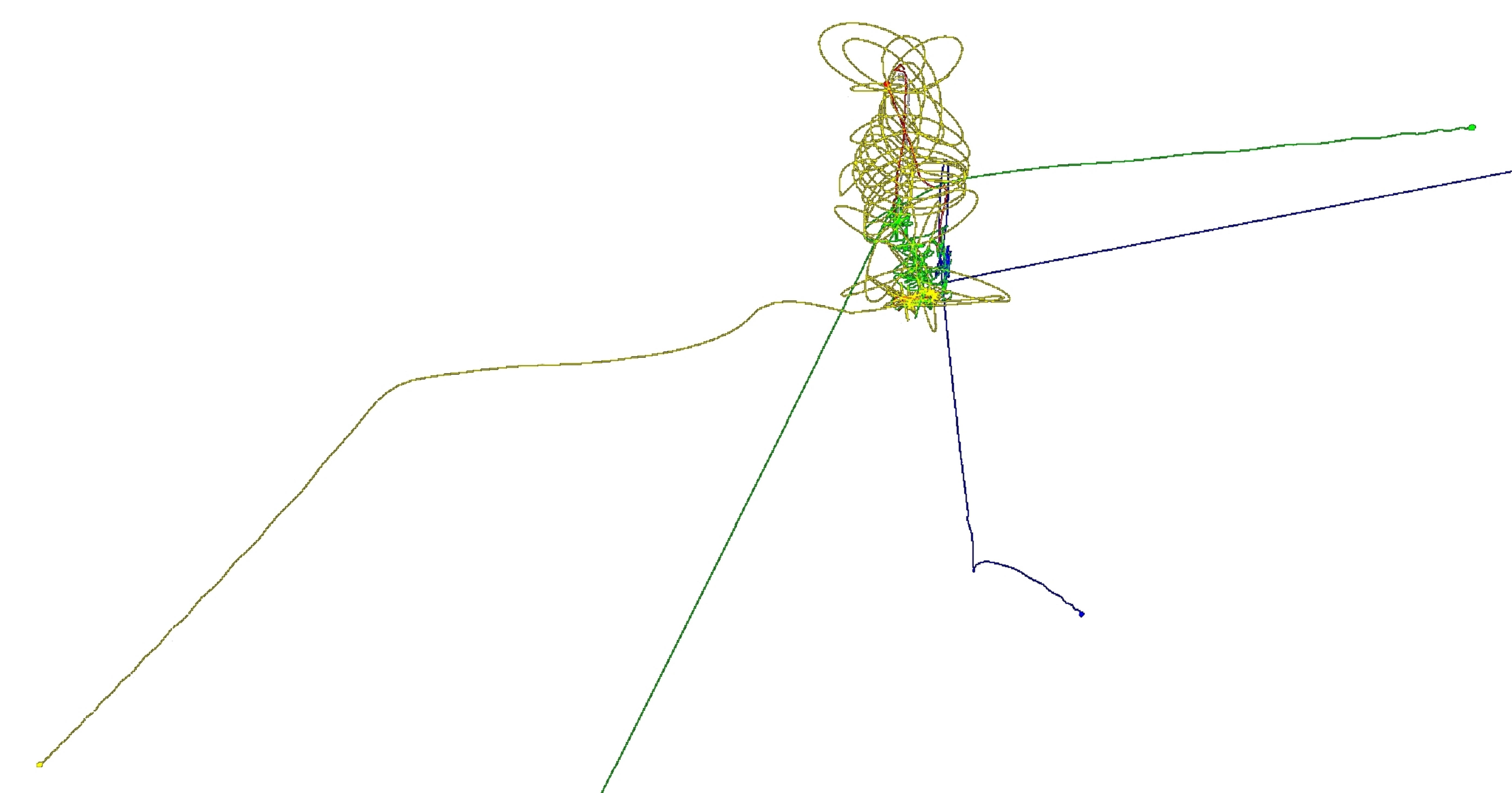}
\end{center}
\caption{
Presentation of the complicated orbital dance of 3 of the minor black
holes with the major black hole (red) within a volume of about
10\,parsec from the 11 merging galaxies of Fig.\,\ref{fig:t0} and
using the highest precision calculation.
\label{fig:BHOrbits}}
\end{figure}

We start with one major galaxy (parent) of mass $M_{\rm parent} =
2.2 \times 10^{12}$\,\Msun\, using $N=2.2 \times 10^5$ particles of
$10^{7}$\,\MSun, each and one supermassive black hole with a mass of
$10^{9}$\,\MSun\, in its center.  The 10 children are scaled down
copies of the parent and distributed according to a
Plummer \cite{1911MNRAS..71..460P} spherical distribution with a
virial radius of 500,000 pc (see \cite{2013MNRAS.431..767B} for
details).  We ignore the gas content of the Galaxy, stellar evolution
and feedback processes. We motivate this limited scope and reduced
physics by the assertion that much of the dynamical processes in which
we are interested are driven by the gravitational evolution of spiral
structure and not by the interstellar gas.  The total amount of gas is
relatively small $\aplt 15\%$ compared to the total mass in stars, and
it is distributed more homogeneously, and is not expected to drive the
global dynamics of the Milky Way.  For the local black-hole dynamics
however, gas may play an important role, but we consider the galaxies
to be dry. In addition, by limiting our scope we are able to run
larger simulations and achieve a higher spatial and temporal
resolution than otherwise possible.  In Fig.\,\ref{fig:t0} we present
a rendering of this initial configuration.

\begin{figure}[t]
\begin{center}
\includegraphics[width=0.5\textwidth]{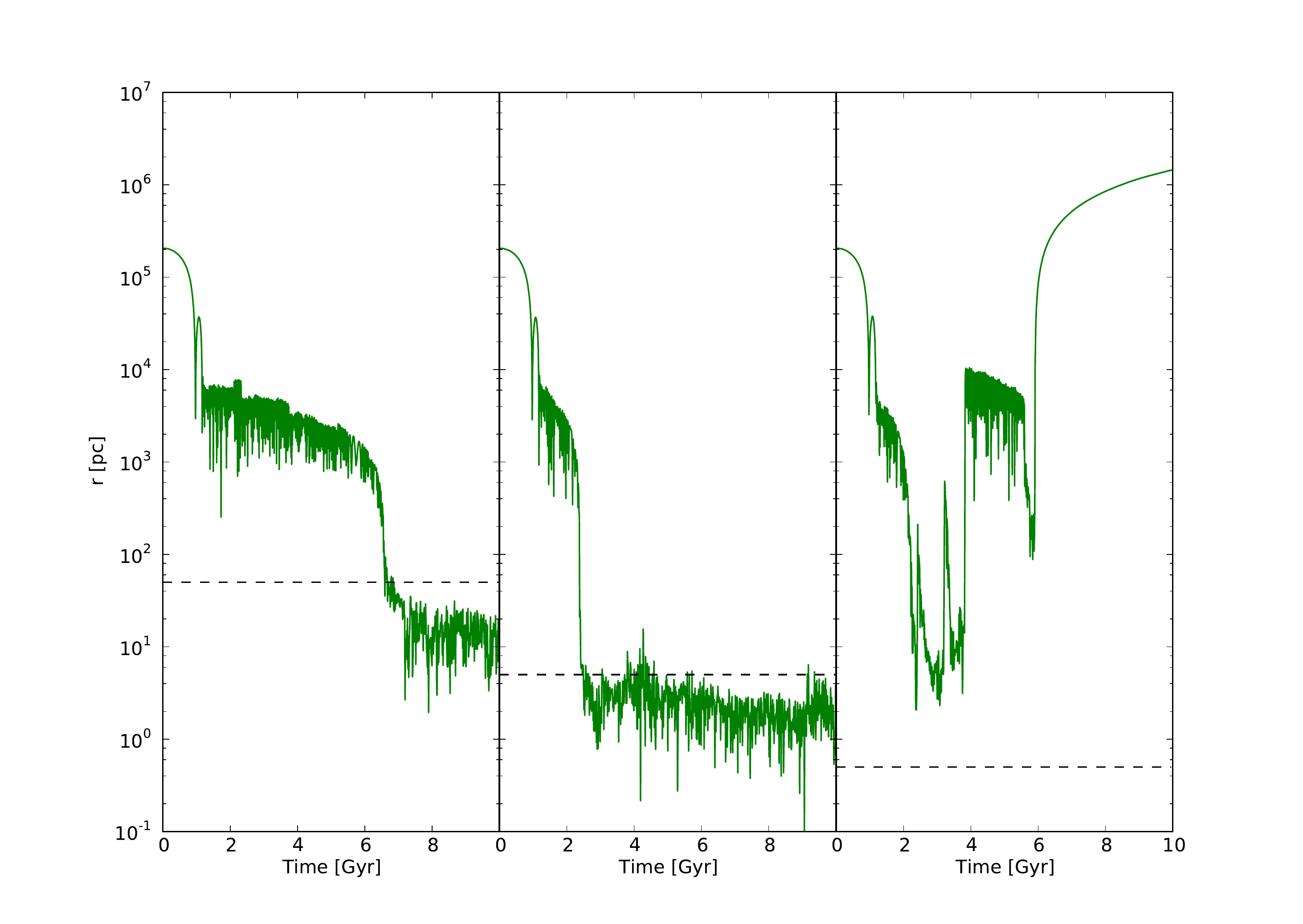}
\end{center}
\caption{
The evolution of the distance between the black hole of the major
galaxy to one of the minor galaxies. In the left panel we adopted a
rather large gravitational softening parameter of 50\,pc (dashed
horizontal line). In the middle panel we adopted 5\,pc softening, and
0.5\,pc in the right most panel.
\label{fig:BHevolution}}
\end{figure}

In Fig.\,\ref{fig:BHevolution} we present the distance from one of the
child's black holes to that of the parent galaxy's black hole.  The
three panels give the result of the same initial realization with a
coupling between the tree and direct $N$-body codes and using different
precision for each of the simulations (parameterized with the integration
time step and the choice of the gravitational softening in the tree
code). In the left most, least accurate simulation one would naively
conclude that the two black holes coalesce about 6\,Gyr after the
start of the simulation. With a more precise (smaller softening length
and time-step) integration (middle panel) the merger appears to occur
at an earlier epoch (at about 2.5\,Gyr). With the most accurate
calculation (right-most image) we see that the incoming black hole is
ejected to large radius after a series of strong encounters between
2\,Gyr and 6\,Gyr. The closest approaches are not resolved in the
figure due to the discrete time-stepping of the output, but the
detailed orbital trajectories of the encounter between several of the
black holes is presented in Fig.\,\ref{fig:BHOrbits}. Of the 10
infalling black holes in the most precise calculation 8 were ejected
and only 2 merged with the parent black hole, whereas in the least
precise calculation all black holes, except one, merged.

In Fig.\,\ref{fig:BHOrbits} we present the orbital evolution of 3 of
the minor black holes and how they interact with the major black hole.
Such complicated orbits cannot be calculated sufficiently accurate
with the tree-code to follow the intricacies of the self-gravitating
system, but require a high-precision direct $N$-body technique to
resolve the subtleties in the chaotic regime, whereas the bulk of the
galaxy material (stars and dark matter) are integrated using the
hierarchical method. One can even question whether or not a direct
method is sufficiently accurate considering the exponential divergence
of the solution, but as has been demonstrated in 
\cite{2014ApJ...785L...3P}, 
the quasi ergodicity of self-gravitating systems allow for relatively
inaccurate calculations to provide statistically meaningful results.

\section{Conclusions}

The use of GPUs has moved high-performance computing to our desktops
and elevated scientific calculations in GPU equipped supercomputers to
new heights. The self-gravitating systems community, which was already
spoiled by the GRAPE family of computers, has been able to quickly
benefit from this relatively new technology due to the similarities in
architecture. Simulations which 5 years ago could only be performed on
expensive special purpose computers or large scale supercomputers can
now be performed on a rather ordinary desktop computer.

The enormous advances in hardware development that has been driving
the computational gravitational dynamics community has managed to
benefit from the use of GPUs by adopting hybrid methods.  These have
enabled simulations with enormous dynamic range and an
unprecedented resolution in mass, and temporal as well as spatial
scales.

With these improvements in hardware the pressure on producing
efficient and accurate software has increased
substantially. Monolithic software cannot cope with the complexities
of real life systems, not in terms of scale (temporal and spatial) and
not in terms of physics (stellar dynamics, hydrodynamics, etc). Novel
numerical techniques are required in order to benefit from the
current hardware. This will also allow us to resolve the scales and
physics required in astronomical applications.

The hybridization of software in order to achieve these objectives is
slowly starting.  At the moment the computational astrophysics
community is driven by these issues.  The Hamiltonian-splitting
strategy (see \S\,\ref{Sect:reunion}) to couple different algorithms
is effective but reduces the overall numerical scheme to second
order. Higher order coupling strategies are in development, but a self
consistent and dynamic coupling between the various scales and physics
in the astronomical applications are still far ahead.

In the calculations we presented here we would like to incorporate
gravitational radiation in order to study what really happens when the
black holes come close together.  This seemingly simple addition to
the physics however, may have dramatic consequences to the overall
integration scheme and therewith to the global structure and the
performance of the implementation.  At this moment we do not know if
the high-accuracy calculation, presented here, gives a reliable answer
to what really happens to the black holes; we still do not know
whether or not the black holes are ejected (as indicated in the
high-precision simulation) or that general relativistic effects causes
them to merge after all.

In a future implementation we consider adding gas via smoothed
particles hydrodynamics, which will also be realized using a
Hamiltonian splitting technique similar to the one described in
section\,\ref{Sect:reunion}. Such a coupling can be realized via the
Astronomical Multipurpose Software
Environment \cite{PortegiesZwart2013456}.

\section*{Acknowledgments}

It is a pleasure to thank Arthur Trew, Richard Kenway and Jack Wells
for arranging direct access via Director's time on ORNL Titan, Mark
Harris, Stephen Jones and Simon Green from NVIDIA, for their
assistance in optimizing {\tt Bonsai} for the NVIDIA K20X, Evghenii
Gaburov, Steve McMillan and Jun Makino for discussions.  Part of our
calculations were performed on Titan (ORNL), Hector (Edinburgh),
HA-PACS (University of Tsukuba), XC30 (National Astronomical
Observatory of Japan) and Little Green Machine (Leiden University).
This work was supported by the Netherlands Research Council NWO
(grants
\#639.073.803 [VICI], \#643.000.802 [JB], \#614.061.608 [AMUSE] and
p\#612.071.305 [LGM]), by the Netherlands Research School for Astronomy
(NOVA).  This research used resources of the Oak Ridge Leadership
Computing Facility at the Oak Ridge National Laboratory, which is
supported by the Office of Science of the U.S. Department of Energy
under Contract No.  DE-AC05-00OR22725.

\section*{The Authors}
\paragraph{Simon Portegies Zwart}
is professor of computational astrophysics at the Sterrewacht Leiden
in the Netherlands.  His principal scientific interests are
high-performance computational astrophysics and the ecology of dense
stallar systems.\\ e-mail:{\tt spz@strw.leidenuniv.nl},\\ 
URL:\url{http://www.strw.leidenuniv.nl/~spz/}

\paragraph{Jeroen B\'edorf}
is a graduate student in the computational astrophysics group of
Portegies Zwart.  His scientific interests are high-performance
computing and graphics processing units. \\ e-mail:{\tt
bedorf@strw.leidenuniv.nl}

\end{document}